# Kinematic Characterization of Micro-Mobility Vehicles During Evasive Maneuvers

Paolo Terranova [a,b] *, Shu-Yuan Liu [c], Sparsh Jain [a,b], Johan Engstrom [c], Miguel Perez [a,b]

[a] *Virginia Tech Transportation Institute, United States*
[b] *Department of Biomedical Engineering and Mechanics, Virginia Polytechnic and State University, United States*
[c] *Waymo LLC, United States*

\* Corresponding author.
E-mail address: pterranova@vt.edu

Keywords:
Micro-mobility vehicles (MMVs)
Personal mobility devices
Vehicle kinematics
Evasive maneuver modeling
Autonomous Vehicle simulations

ABSTRACT

*Introduction*: Over the last decade, the increasing popularity of Micromobility Vehicles (MMVs) has led to profound changes in personal mobility, raising concerns about road safety and public health. As a result, there is an increasing need to effectively capture the kinematic performance of different device types, ultimately leading to a comprehensive characterization of their safety boundaries. Hence, this study aims to (1) characterize the kinematic behaviors of different MMVs during emergency maneuvers, (2) explore the influence of different MMV power sources on the device performances, and (3) investigate if a piecewise linear model is suitable for modeling MMV trajectories. *Method*: A test track experiment where 40 *frequent* riders—riding a subset of two different types of electric MMVs (two handlebar-based and two handlebar-less), their traditional counterparts, and, in some cases, behaving as running pedestrians—*performed* emergency braking and swerving maneuvers. A second experiment was conducted to determine the minimum radius of curvature, and thus the swerving boundary of different MMVs. *Results*: Device power source *has a statistically significant influence on* kinematic capabilities of the MMVs. While electric MMVs (e-MMVs) displayed superior braking capabilities compared to their traditional counterparts, the opposite was observed in terms of swerving performance, with traditional devices outperforming electric devices. Furthermore, performance varied significantly across the different MMV typologies, with handlebar-based devices (i.e., bicycles and scooters) consistently outperforming the handlebar-less devices (i.e., skateboards and onewheel) across nearly all the metrics considered. The piecewise linear models used for braking profiles were found to fit well for most MMVs, except for skateboards and pedestrians where foot *engages with the ground during* maneuvers. *Conclusions*: This research highlights the influence of vehicle-specific characteristics on the stability, agility, and maneuvering capabilities of the different MMV types. There are significant differences between electric-powered and their traditional counterparts, as well as between handlebar-based and handlebar-less models. These findings underscore that the effectiveness of steering or braking in preventing collisions may vary depending on the type and power of the device. This study also demonstrates the applicability of piecewise linear models for generating parameterized functions that accurately model braking trajectories, providing a valuable resource for traffic event reconstructions, simulations, and for automated systems developers. The model, however, also reveals that the single brake ramp assumption does not apply for certain types of MMVs or for pedestrians, indicating the necessity for further improvements.

# 1. Introduction

Micromobility vehicles (MMVs), as defined in the SAE J3194 standard (2019), have led to significant changes in personal mobility over the last decade (Zagorskas & Burinskiene, 2020). Primary types of MMVs are bicycles and e-scooters, which in 2019 facilitated more than 136 million trips through over 190 shared mobility programs in the United States (GHSA, 2020; NACTO, 2020). Additionally, there are other MMV types—such as powered self-balancing (e.g., onewheel, Segway, monocycle) and powered non-self-balancing (e.g., e-skateboard)—which are less widespread partially due to their novelty in the market. All these MMVs are generally employed for daylight, clear-weather, and relatively short (< 5 km) trips, which nonetheless account for a significant proportion of all personal mobility trips (Abduljabbar et al., 2021). Moreover, MMVs are increasingly substituting both walking and driving trips as a faster, sustainable, and cost-effective means of transportation (Chang et al., 2019; GHSA, 2020; NACTO, 2020). However, despite being recognized as a potential solution to reduce pollution and congestion of urban environments (Abduljabbar et al., 2021; Smith & Schwieterman, 2018), these new personal mobility devices represent a new source of challenges for road safety and public health (Zagorskas & Burinskiene, 2020).

Specifically, MMV riders have been observed to not follow traffic rules, raising concerns about a potentially elevated frequency of conflicts with other road users (Dozza et al., 2016; Haustein & Møller, 2016; Langford et al., 2015). This behavior has contributed to the rising and significant number of MMV-related injuries (B. Trivedi et al., 2019). In the United States, for instance, analysis of medical records has revealed a substantial increase in MMV-related traumas since the launch of sharing programs for e-scooters (Badeau et al., 2019; T. K. Trivedi et al., 2019), e-bicycles (Cicchino et al., 2021), and Segways (Boniface et al., 2011). Similar findings have also been reported in China (Zhang et al., 2015), Europe (Berk et al., 2022; Schepers et al., 2014; Stigson et al., 2021), and Australia (Mitchell et al., 2019). Additionally, despite the significant prevalence and severity of head and facial injuries in MMV riders (T. K. Trivedi et al., 2019; Wüster et al., 2021), the usage of helmets remained low (Badeau et al., 2019; Moftakhar et al., 2021; T. K. Trivedi et al., 2019). Furthermore, differences in injury patterns have been observed among different MMV types, with electric-powered MMVs (e-MMVs) generally found to have higher crash risk and more severe injury outcomes than traditional MMVs, a phenomenon potentially connected with the higher prevalence of older e-MMV riders (Berk et al., 2022; Cicchino et al., 2021; Haustein & Møller, 2016) and their higher operational speeds compared to traditional MMVs (Schepers et al., 2014; Schleinitz et al., 2017). In addition to this injury trend, even the integration of MMVs into the current transportation systems is complicated by the substantial variability in riding behaviors and kinematic capabilities among different MMV types (Dozza et al., 2016, 2023; Vetturi et al., 2023), and by the new interactions they establish with other road users (James et al., 2019; Pashkevich et al., 2022; Petzoldt et al., 2017; Stigson et al., 2021; Tark, 2020). The characterization of the kinematic capabilities of different MMV types could contribute to a better understanding of their collision avoidance behaviors and physical limits, allowing for the development or improvement of effective countermeasures such as Advanced Driver Assistant Systems (ADAS) or Advanced Driving Systems (ADS). Unfortunately, this characterization is constrained by the current limitations of suitable data. Retrospective crash data are limited to incident descriptions such as police reports (Shah et al., 2021; Weber et al., 2014), hospital records (Badeau et al., 2019; Berk et al., 2022; Schepers et al., 2014), or newspapers (Yang et al., 2020). In the same way, naturalistic studies on MMVs (e.g., Pashkevich et al., 2022; Todd et al., 2019; Vlakveld et al., 2015; White et al., 2023) have primarily focused on investigating MMV riding behaviors and crash factors without extensively exploring the kinematic aspects of the devices. While this type of data can be useful for developing strategies and recommendations, it is insufficient to fully understand and reconstruct, and ultimately prevent, specific crash circumstances.

A novel framework by Dozza et al. (2022) was proposed for the experimental collection of MMV longitudinal and lateral dynamics. Though their methodology provides a useful approach for MMV kinematic characterization, the generalizability of their findings is constrained by the limited number of participants and of the different devices employed. Similarly, Vetturi et al. (2023) estimated and modeled bicycle and e-scooter braking distances in planned maneuvers, with a limited number of devices and participants (N = 3), finding no significant differences between the two MMVs. In a different track experiment, Dozza et al. (2023) characterized the MMV longitudinal kinematics of electric and conventional bicycles, e-scooters, and Segways, modeling both planned and unplanned braking maneuvers with linear regression techniques. Their results highlighted significant differences between traditional and e-bicycle performance, emphasizing the need to deeply understand the potential distinctions among devices of the same type but with varying power sources. Additionally, the findings also indicated that the Segway displayed the poorest performance when compared to all the other devices. However, the experimental design lacked controls for participants' familiarity with the devices, leading to a substantial amount of missing data for the Segway (approximately 50%) and limiting the applicability of the ensuing model.

With respect to characterization of MMV lateral dynamics, Garman et al. (2020) and Lee et al. (2020) initiated investigations into the performance of e-scooters and bicycles. Their results, however, are limited to these two devices and did not involve emergency maneuvers. Additionally, in a recent study conducted by Li et al., (2023), the authors explored MMV collision avoidance performance in longitudinal and lateral maneuvers, modeling the latter performances using an arctangent model. Their experiment considered e- and traditional bicycles and two different e-scooters, where participants were asked to brake and steer to avoid a fixed soft cardboard car placed on their path. While their findings offer valuable insights into testing and modeling techniques for MMVs, there are still unanswered questions regarding the performance of a broader range of devices. Furthermore, in terms of lateral metrics, the authors restricted their analysis to lateral offset and steering angle, neglecting lateral acceleration and jerk, important measures to fully evaluate steering performance (Brännström et al., 2014; Kovácsová et al., 2016). Additionally, the authors noticed a slight alteration in participant behavior, as riders tended to make minor adjustments by steering away or slowing down just before commencing the actual maneuver due to their awareness of the obstacle's presence.

Motivated by these gaps, three primary research questions were examined in this study. (1) Do MMV types and running pedestrians exhibit different longitudinal and lateral performances during evasive maneuvers? (2) Do these differences, if present, vary as a function of the MMV propulsion system (i.e., electric and traditional)? (3) Is a piecewise linear model previously applied to passenger vehicles and trucks (Markkula et al., 2016), suitable to model MMVs' braking and running pedestrians' stopping trajectories? The main hypotheses were that different MMV types and propulsion systems would influence the MMVs' longitudinal and lateral evasive maneuvers, and that the piecewise linear model could be used to effectively model MMV braking trajectories.

## 2. Method

This test track experiment used a mixed-factor design split by Maneuver (*Braking* or *Swerving*) while blocking for Sex (*Male* or *Female*, when allowed by the participant pool) and Speed (*Low* or *High*, see **Table 1**), and including MMV Type (*Onewheel, Bicycle, Scooter, Skateboard,* or *Pedestrian*) and MMV Power Mode (*Electric* or *Traditional*, when applicable) as main effects. MMV Type was a between-subjects factor, whereas MMV Power was a within-subject factor. The MMVs were ridden by self-reported experienced participants on a closed test-track course. Participants completed braking and swerving maneuvers at different speeds while kinematic and video data were recorded.

### 2.1. Participants

A total of 40 participants (27 males and 13 females) were included in the study, with a mean age of 28.3 years (SD = 10.8, median = 22.5). There was an intent to balance participant sex for each MMV, but this proved to be challenging for the onewheel and skateboards, where candidate participants who responded to the study advertisements were predominantly male. Candidates were determined eligible for a particular MMV if they reported using that MMV weekly or more frequently. Ten out of the 40 participants also completed trials running as pedestrians. Each participant had to pass vision and hearing screenings. A weight limit of 220 lbs. was established to comply with the device manufacturers' weight recommendations. Female participants were also given the option to take a pregnancy test, since pregnant individuals were ineligible to participate due to potential fall risk. To increase safety, participants were also asked to wear long-sleeve shirts and long pants, close-toed shoes, helmets, and elbow and knee pads throughout the experiment. The experiment was approved by the Virginia Tech Institutional Review Board. All participants provided informed consent.

## 2.2. Equipment

The MMVs used in this study were (**Figure 1**):
- *Bicycle*: E-bicycle – Vivibike C26 250W City cruiser; Traditional bicycle– Trek FX 7.2 - 24 inch
- *Scooter*: E-scooter – Ninebot KickScooter G30P; Traditional scooter – Swagtron K8 Titan
- *Skateboard*: E-skateboard – Caroma37 Inch 350W E-Skateboard; Traditional skateboard – Element Seal Skateboard; Traditional longboard – Apollo Longboard Bali
- *Onewheel*: Future Motion Onewheel Pint

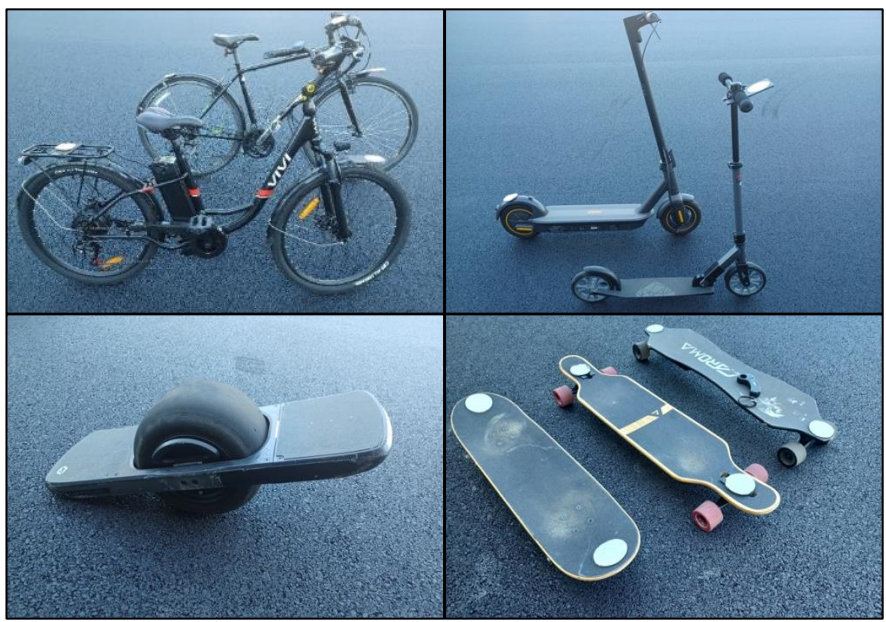

*Figure 1. MMVs used in the experimental trials, clockwise from top left: e- and traditional bicycles, e- and traditional scooter, e-skateboard, traditional longboard, and traditional skateboard, and onewheel.*

Participants were assigned to the onewheel, e-bicycle, e-scooter, or e-skateboard MMVs based on their use experience. For all MMVs except the onewheel, participants assigned to the electric version of the MMV were also asked to perform the same tests with the corresponding traditional version(s) of the MMV. Participants selected to serve as pedestrians were also asked to complete the same maneuvers while walking/jogging. Data was collected using a combination of overhead camera recordings (Alied MAKO G-319C with a Computar a4z2812cs lens) and an inertial measurement unit (IMU) sensor (APDM Opal V2R). A retroreflective vinyl marker was placed atop the participant's helmet to enhance visibility of this feature in the videos, which were recorded at 30 Hz. The IMU was secured in the participant's lumbar area using an adjustable Velcro strap. IMU data was recorded at 128 Hz.

## 2.3. Protocol

To minimize adaptation effects, participants began with "warm-up" familiarization trials for each new set of conditions, including changes in MMV type, speed, or maneuver. In these trials, participants practiced reaching cruising speed, swerving, and braking before a landmark. Participants were allowed to repeat familiarization trials as often as they wanted before engaging in actual experimental trials, which consisted of braking and swerving maneuvers.

### 2.3.1. *Braking and swerving maneuvers*

In the braking maneuver (**Figure 2** top), participants were asked to reach a prescribed cruising speed, then, after a red light illuminated on a traffic signal in the distance, they had to perform a braking maneuver as hard as possible while maintaining control. In the swerving maneuver (**Figure 2** bottom), participants were asked to follow a center line on the pavement at the cruising speed. Then, upon an indication by a directional (i.e., left or right) arrow signal, participants had to swerve in the direction indicated by the arrow, as quickly as possible, while staying within demarcated lateral boundaries and maintaining control of the MMV. The order of directional trials was randomized following pre-generated treatment order sheets to reduce anticipation effects.

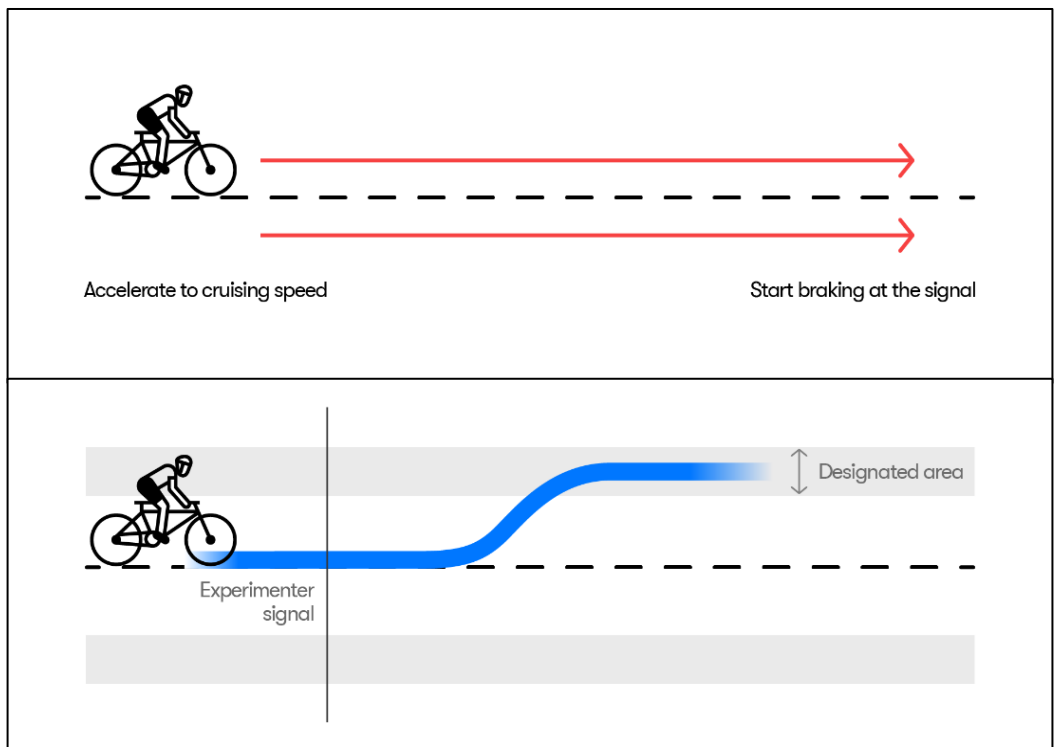

*Figure 2. Braking (top) and swerving (bottom) maneuvers.*

The order of the maneuvers (braking or swerving) and the order of the MMVs (electric or traditional) for each participant were randomized following pre-generated orders. All participants, however, completed the maneuvers first at a "low" speed and then at the "high" speed (**Table 1**). Trial repetitions were required until participants were able to reach the required speed range for each maneuver. The direction of swerving was also randomized for each trial of each MMV. After a successful braking or swerving maneuver, participants were asked to verbally rate their perceived effort during the previous maneuver (e.g., for braking. "*On a scale of 1 to 10, with 10 being the hardest you think you can stop [using this device while maintaining control of the device]/[while running], how hard do you think you stopped in this last trial?*"). If the score provided was under seven (a threshold that was not shared with participants), participants were requested to repeat the trial, up to five times. Left or right swerve trials were randomized regardless of the score and continued in such fashion until participants had provided scores greater than or equal to seven for both left and right swerve directions.

*Table 1. Target and speed ranges for each MMV in the study.*

| MMV Type | Speed | | |
|---|---|---|---|
| | Speed Condition | Target in m/s (Target in mph) | Range in m/s (Range in mph) |
| Traditional Bicycle (T-bicycle) | Low | 2.9 (6.5) | 2.7–3.6 (6.0–8.1) |
| | High | 4.5 (10.1) | 4.0–4.9 (8.9–11.0) |
| Electric Bicycle (E-bicycle) | Low | 3.4 (7.6) | 2.7–3.6 (6.0–8.1) |
| | High | 4.5 (10.1) | 4.0–4.9 (8.9–11.0) |
| Traditional Scooter (T-scooter) | Low | 3.1 (6.9) | 2.7–3.6 (6.0–8.1) |
| | High | 4.0 (8.9) | 4.5–5.4 (10.1–12.1) |
| Electric Scooter (E-scooter) | Low | 3.1 (6.9) | 2.7–3.6 (6.0–8.1) |
| | High | 5.8 (13.0) | 4.5–5.4 (10.1–12.1) |
| Traditional Skateboard (T-skateboard) | Low | 3.1 (6.9) | 2.2–3.1 (4.9–6.9) |
| | High | 4.0 (8.9) | 3.8–4.7 (8.5–10.5) |
| Electric Skateboard (E-skateboard) | Low | 2.5 (5.6) | 2.2–3.1 (4.9–6.9) |
| | High | 4.0 (8.9) | 3.8–4.7 (8.5–10.5) |
| Traditional Longboard (T-longboard) | Low | 2.7 (6.0) | 2.2–3.1 (4.9–6.9) |
| | High | 4.7 (10.5) | 3.8–4.7 (8.5–10.5) |
| Onewheel | Low | 2.9 (6.5) | 2.5–3.4 (5.6–7.6) |
| | High | 4.5 (10.1) | 4.0–4.9 (8.9–11.0) |
| Pedestrian | Low | 3.4 (7.6) | 2.9–3.4 (6.5–7.6) |
| | High | 6.0 (13.4) | 5.6–6.5 (12.5–14.5) |

### 2.3.2. *Radius of curvature*

In order to determine the swerving limits of each device, a separate experiment was completed to determine the minimum radius of curvature (the minimum curvature at the lowest speed) each MMV was able to attain. A total of three expert participants per type of MMV were recruited and asked to ride the applicable MMVs inside the circular areas illustrated in **Figure 3**.

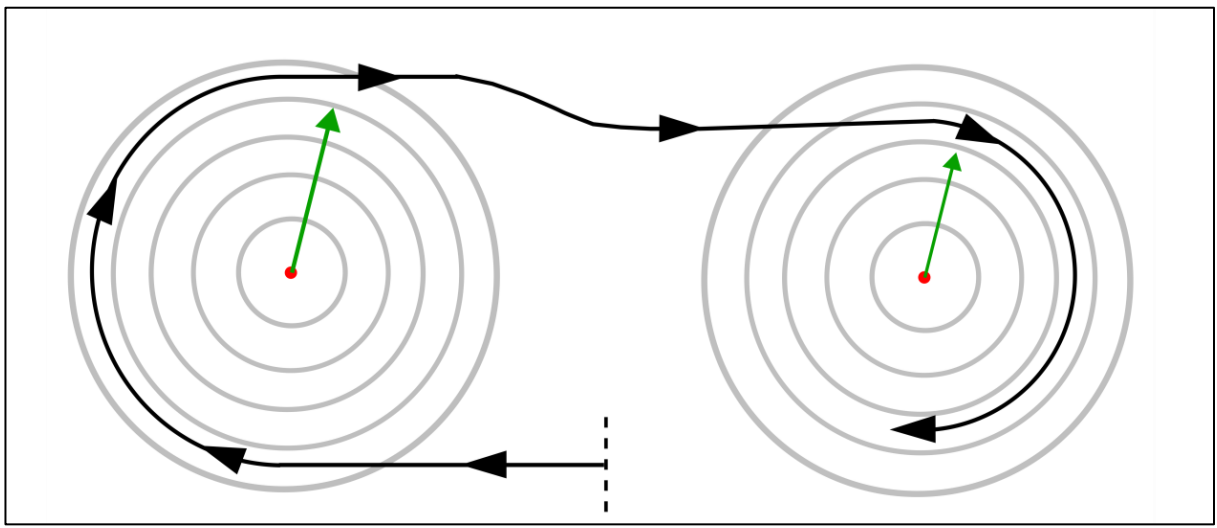

*Figure 3. Circular areas used to obtain the minimum radius of curvature.*

Participants started at the dashed line in **Figure 3** and made the first turn circling inside the designated area between the two largest diameter circles. Once they traversed 180° inside the initial designated area, participants followed a mostly straight path towards the opposite set of circles. While in this section, participants adjusted their speed, roll, and steering angle, as necessary, to prepare themselves to repeat the maneuver between the next two smaller diameter circles in the opposite set of circles. This procedure was repeated, on opposite circles and traversing smaller diameter regions, until the participant either had to put a foot on the ground to maintain balance or was not able to perform the full 180° turn inside the designated area. Participants were asked to perform the maneuver while maintaining the minimum possible speed that allowed them to maintain balance and traverse the smallest-diameter circular area possible.

## 2.4. Data Analysis

Calibration tests provided the camera's intrinsic and extrinsic parameters, enabling the fusion of relevant track views into a continuous video for precise translation between pixel locations and (X, Y) ground plane coordinates. The resultant videos were coded through a human-supervised automated algorithm to track the coordinates of the participant's head marker during the maneuver. Head position data was filtered with a third order zero-lag Butterworth low-pass filter with a cutoff frequency of 1 Hz to provide a processed signal. This processed signal was then differentiated and filtered again to provide a speed signal. IMU data was filtered using a sixth order zero-lag Butterworth filter with a cutoff frequency of 10 Hz and the sensor orientation was corrected for any leaning of the participants. The fully processed IMU data provided longitudinal, lateral, and vertical accelerations in the upright MMV's frame. The speed signals from cameras and the acceleration data from the IMU were synchronized to a common clock, which was arbitrarily assigned to be zero when the light indicating that participants should stop or to swerve in a certain direction illuminated. An over filtered version of the IMU data (sixth order zero-lag Butterworth filter with a cutoff frequency of 1 Hz) was also created and used to identify the maneuver start and end times, and the other time-relevant points reported in **Table 2** and illustrated in **Figure 4**.

*Table 2: Time-relevant metrics used to identify the different regions of each maneuver*

| Time-relevant points names | Operational definitions |
|---|---|
| Maneuver Start Time | Time when the over filtered acceleration first exceeded 0.25 m/s$^2$ after light activation |
| Peak Acceleration Time | Time of the first maximum/minimum peak of the over filtered acceleration |
| Acceleration Inflection Time* | Time when the over filtered acceleration reached zero the second time after light activation |
| Acceleration Secondary Peak Time* | Time of the second maximum/minimum peak of the over filtered acceleration |
| Maneuver End Time | Time when longitudinal or lateral speed either became zero or stabilized close to that value |
| * Only calculated for swerving maneuvers | |

The position, speed, acceleration, and the time-relevant points were used to isolate the relevant regions in the maneuver kinematics. During the braking trials, a single region spanning from the start to the end of the maneuver was identified. In contrast, two distinct regions were identified in swerving maneuvers: the initial segment spanned from the maneuver start to the inflection time, representing participants steering to enter the designated area; the second segment extended from the inflection to the end time, representing participants steering to maintain the trajectory within the designated area (**Figure 4**).

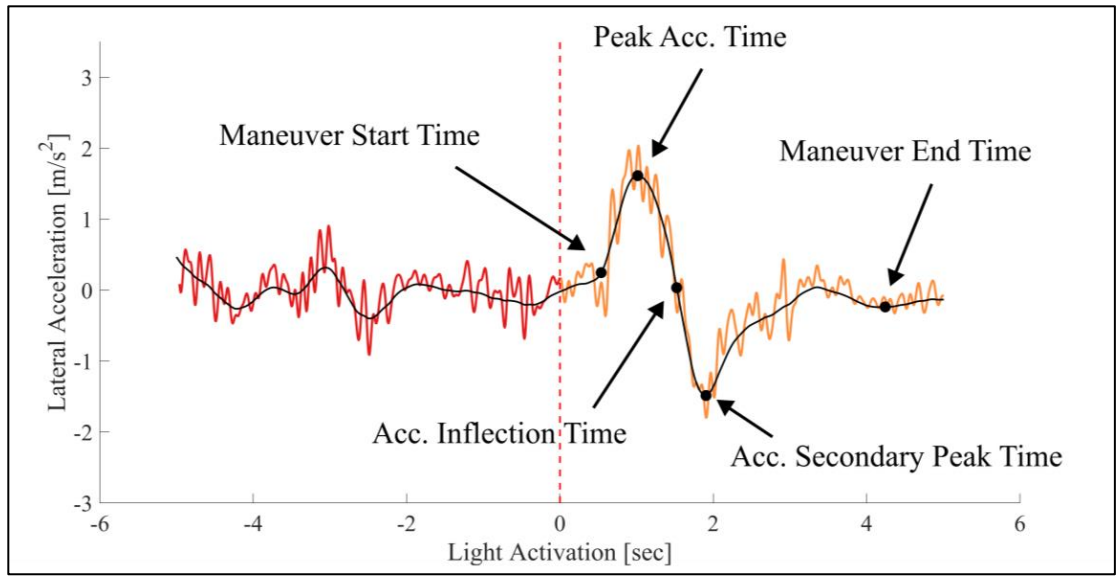

*Figure 4. Illustration of time points on a sample lateral acceleration signal trace. The smooth curve, on which the timepoints are indicated, depicts over-filtered acceleration data provided to show the general acceleration shape. The jagged curve depicts the corresponding filtered IMU acceleration.*

Once the relevant regions for each maneuver were identified, the metrics reported in **Table 3** were calculated for each maneuver. Notably, the differentiation between the two regions of swerving maneuvers led to the formulation of a more extensive range of metrics for the swerving trials. This extended set includes two average deceleration values for the two maneuver regions and three average jerk values linked to the three ramps of the maneuver.

*Table 3: Names and operational definition of the metrics of interest*

| Metrics names | Operational definitions |
|---|---|
| Mean Long. Acc. | Arithmetic mean of the longitudinal acceleration sampled between the *maneuver start time* and *maneuver end time* |
| Mean Long. Jerk | Mean jerk (Δ acceleration/time elapsed) observed between the *maneuver start time* and the *peak acceleration time*. |
| Mean Lat. Acc. Initial Maneuver* | Arithmetic mean of the lateral acceleration sampled between the *maneuver start time* and *acceleration inflection time*. |
| Mean Lat. Acc. Rebound* | Arithmetic mean of lateral acceleration sampled between the *acceleration inflection time* and the *maneuver end time*. |
| Mean Lat. Jerk Onset to Peak Acc.* | Mean jerk (Δ acceleration/time elapsed) observed between the *maneuver start time* and the *peak acceleration time*. |
| Mean Lat. Jerk Peak Acc. to Rebound Peak* | Mean jerk (Δ acceleration/time elapsed) observed between the *peak acceleration time* and the *acceleration secondary peak time*. |
| Mean Lat. Jerk Rebound Peak to End* | Mean jerk (Δ acceleration/time elapsed) observed between the *acceleration secondary peak time* and the *maneuver end time*. |
| * Absolute values were considered to allow comparison across the swerve directions | |

The minimum radius of curvature was estimated averaging the smallest radius participants could attain at the lowest speed, while maintaining the MMVs control. The minimum speed for these maneuvers was extracted from the video recordings.

Generalized linear mixed-effect models were used to assess the significance of the MMV Type (eight different MMV types and the pedestrian) and MMV Power Source factors (Electric vs Traditional) on the relevant metrics calculated from the data (**Table 3**), controlling for participant sex and maneuver speed. The MMV Type influence was assessed considering all the devices and the pedestrians. When analyzing the influence of MMV Power, however, pedestrians and the Onewheel were excluded due to the lack of comparable counterparts. *Post-hoc* Tukey honestly significant difference tests were used to assess the statistically significant treatment levels within statistically significant factors. Statistical significance was assessed using a Type I error (α) of 0.05 for the MMV Type analysis, and 0.10 for the MMV Power Source analysis (due to the reduced sample size for the analysis).

### 2.4.1. *Braking profile modeling with piecewise linear model*

To enable trajectory reconstructions, the braking profiles were modeled using piecewise linear models. This approach has been previously used to model evasive braking of passenger vehicles and trucks, and showed proper fit (Markkula et al., 2016). A piecewise linear model consists of the following parameters and is illustrated in **Figure 5**:

- The initial steady state acceleration, $a_0$, which can be non-zero, and must be greater than the final steady state, $a_1$;
- The time of braking, $t_b$, at which the acceleration ramps down to the final steady state acceleration at a rate of the jerk, $j_b$;
- The braking jerk, $j_b$, at which the agent ramps down to the final steady state; and
- The final steady state acceleration, $a_1$, which is negative.

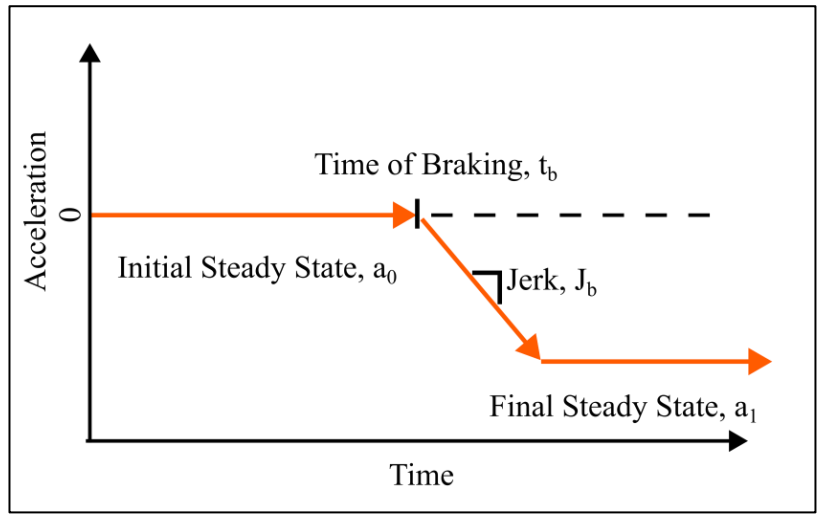

*Figure 5. An illustration of a piecewise linear model.*

The longitudinal acceleration data from fully processed IMU data was additionally filtered with a second order zero-phase digital filter with cutoff frequency of 3 HZ. The purpose of the filter was to remove excessive noise that can interfere with model fitting, while retaining key kinematic characteristics of the maneuver. For each braking trial, the piecewise linear model fitting was done between 0.5s before Maneuver Start Time and when the acceleration first crossed a pre-specified threshold after the Peak Acceleration Time (see **Table 2** The thresholds varied by MMV types due to device-wise kinematic differences. Thresholds for bicycles, traditional scooters, e-scooters, onewheels, skateboards, and pedestrians were –2.5m/s$^2$, -1.5m/s$^2$, -2.5 m/s$^2$, -1.5 m/s$^2$, and –0.5 m/s$^2$, respectively. All events were manually reviewed to ensure the start and end points indeed encompassed the entire braking duration needed for the model.

Generalized linear mixed-effect models supplemented by *post-hoc* Tukey honestly significant difference tests (α=0.05) were employed to assess the efficacy of the model in capturing variations in braking performance across diverse MMVs. This analysis was performed to allow the comparison of the model-generated outcomes with the corresponding experimental results.

## 3. Results

The 40 participants recruited in the study performed a total of 540 trials. Out of these, 14 were discarded due to video recording failures. An additional 13 trials had to be discarded due to lack of accelerometer data or insufficient information to properly synchronize the accelerometer and the video data. These exclusions left a total of 513 maneuvers for analysis.

### 3.1. Braking Trials

#### 3.1.1. *Experiment results and statistical analysis*

The experimental findings revealed that the E-bicycle, T-bicycle, and E-scooter exhibited comparable braking acceleration (Mean Long. Acc.), yielding significantly higher values than the remaining devices **(Table 4)**. In contrast, the T-scooter emerged as the only handlebar-based device showing an inferior acceleration performance, which instead was similar to the handlebar-less skateboard devices. Specifically, the skateboard devices examined showed nearly identical deceleration patterns but noteworthy variations in jerk, with the longboard exhibiting larger mean jerk than other skateboards (albeit not to a statistically significant extent). The last handlebar-less device examined, the onewheel, was able to achieve above-average (but gradual, based on the low mean jerk) acceleration values. In contrast, pedestrian mean jerk was higher than for all other devices, likely because of the high-impulse contact between participants' shoes and the ground required to come to a stop. The mean jerk metric for pedestrians also showed the highest standard deviation amongst all MMV Types.

*Table 4: Descriptive statistics (Mean ± Standard Deviation) and post-hoc test results for the MMV Type factor during braking trials. Devices not connected by the same bracketed letters were significantly different (α=0.05).*

|  | Mean Long. Acc. [m/s$^2$] | Mean Long. Jerk [m/s$^3$] |
|---|---|---|
| E-bicycle | -2.6±0.6 [A] | -6.5±2.2 [B] |
| T-bicycle | -2.5±0.7 [A] | -7.7±3.3 [B] |
| E-scooter | -2.8±0.5 [A] | -6.8±2.3 [B] |
| T-scooter | -1.2±0.3 [C] | -1.5±2.0 [B] |
| E-skateboard | -1.4±0.5 [B,C] | -6.3±4.4 [B] |
| T-skateboard | -1.4±0.3 [B,C] | -6.9±8.7 [B] |
| T-longboard | -1.4±0.3 [B,C] | -9.9±14.0 [B] |
| Onewheel | -1.9±0.3 [B] | -3.4±1.7 [B] |
| Pedestrian | -1.7±0.7 [B,C] | -26.0±26.2 [A] |

Different MMV power sources significantly influenced some devices' braking capabilities, with MMVs powered by electric propulsion showing significantly superior performance compared to their traditional counterparts for bicycles and scooters **(Table 5)**. No statistically significant differences due to propulsion system, however, were observed in braking capabilities between traditional- and e-skateboards.

*Table 5: Post-hoc test results for the MMV Power Source factor during braking trials.*

|  | **Significant Comparisons** |
|---|---|
| Mean Long. Acc. | E-bicycle > T-bicycle *<br>E-scooter > T-scooter *** |
| Mean Long. Jerk | E-scooter > T-scooter *** |
| ***α=0.01; **α=0.05; *α=0.10 | |

### 3.1.2. *Braking profile modeling with piecewise linear models*

One implicit assumption of a piecewise linear model that was not explicitly discussed in the literature is that evasive braking maneuvers must consist of only a single braking action. This is true for human drivers operating a passenger vehicle or truck according to Markkula et al. (2016). Likewise, in the current application, the model fit well for bicycles, Onewheel (**Figure 6**), and scooters. For those devices, stopping or slowing down only requires users to brake once. The assumption, however, did not hold for pedestrians and skateboards in this investigation. Specifically, pedestrians usually needed several consecutive and typically heterogeneous steps until fully stopped (see **Figure 7**).

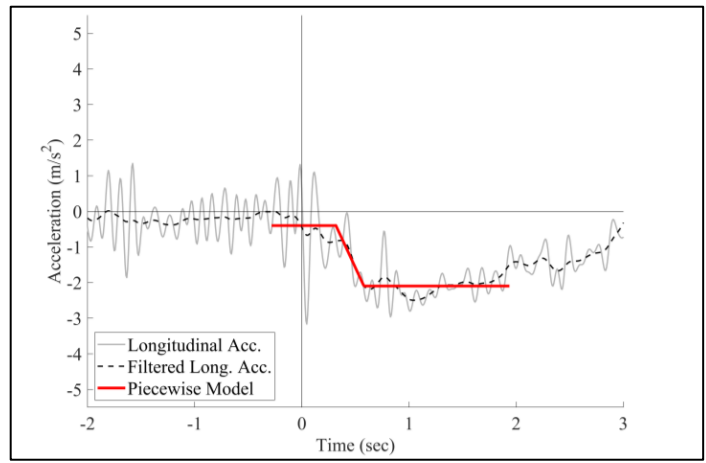

*Figure 6. An example of piecewise linear model fit of a onewheel braking maneuver.*

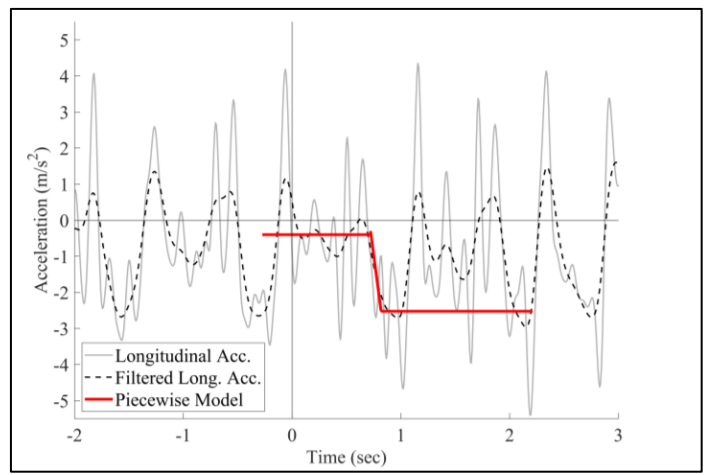

*Figure 7. An example when the piecewise linear model failed to achieve proper fit - a runner braking maneuver.*

Similarly, skateboard riders usually need to drag their feet on the ground more than once to make a full stop, which resulted in several either homogeneous (more common) or heterogeneous (less common) braking ramps in one braking trial (see **Figure 8**). Braking on e-skateboards can be controlled completely by the electric controller with no foot engagement required, which would present as a single braking ramp that fits the piecewise linear model; however, we only observed 3 out of 19 valid trials where participants only used the controller to brake. Multiple braking slopes can make the piecewise linear model completely fail (see **Figure 8**, red line), overestimating or underestimating the jerk when the model averages out the final steady state braking (see **Figure 8**, step in red line).

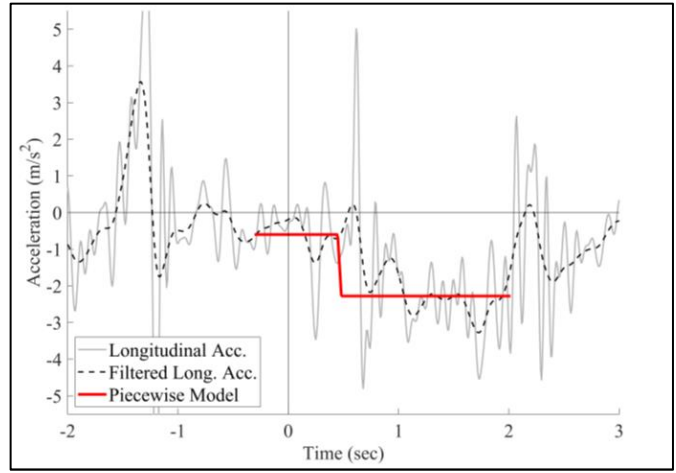

*Figure 8. An example when the piecewise linear model failed to achieve proper fit – a skateboard braking maneuver.*

Therefore, due to these multiple "braking ramps", piecewise linear models were not used to model braking performances for pedestrians nor the three types of skateboards. Model results were summarized using the mean and the standard deviation of the model-derived parameters for the remaining MMV Types **(Table 6)**.

*Table 6: Descriptive statistics (Mean ± Standard Deviation) of final steady state acceleration and jerk from the best fit piecewise linear model.*

|  | Mean of final steady state acceleration [$m/s^2$] | Mean of jerk [$m/s^3$] |
|---|---|---|
| E-bicycle | -3.6±0.6 | -17.2±6.6 |
| T-bicycle | -3.9±0.7 | -20.6±7.0 |
| E-scooter | -4.3±0.6 | -17.0±11.1 |
| T-scooter | -1.8±0.3 | -14.8±12.2 |
| Onewheel | -2.5±0.5 | -7.2±2.4 |

Statistical comparison between the model and the experimental results supported the model's effectiveness in capturing braking performance differences between different MMVs (**Table 7**). The model effectively captured longitudinal acceleration trends across the different MMVs, except for a significant difference between E-bicycle and E-scooter that was not observed in the experimental results. Similarly, for longitudinal jerk, the model highlighted the same pattern of significance as the experimental results with the exception of the T-scooter, which was statistically similar to the other handlebar-based MMVs according to the model results.

*Table 7: Post-hoc test results comparing model-derived and experimental acceleration and jerk during braking trials. Devices not connected by the same letters were significantly different (α=0.05).*

| Model | Longitudinal Acceleration | | | | Longitudinal Jerk | |
|---|---|---|---|---|---|---|
| T-scooter | A | | | | A | B |
| Onewheel | | B | | | A | |
| E-bicycle | | | C | | | B |
| T-bicycle | | | C | D | | B |
| E-scooter | | | | D | | B |
| **Experiment** | **Longitudinal Acceleration** | | | | **Longitudinal Jerk** | |
| T-scooter | A | | | -- | A | |
| Onewheel | | B | | -- | A | |
| E-bicycle | | | C | -- | | B |
| T-bicycle | | | C | -- | | B |
| E-scooter | | | C | -- | | B |

## 3.2. Swerving Trials

### 3.2.1. *Experiment results and statistical analysis*

The experimental findings for the swerving maneuvers showed a greater ability of pedestrians to reach higher acceleration and jerk values compared to the MMVs examined (**Table 8**). Interestingly, the only MMV with statistically comparable swerving performance to the pedestrian was the T-scooter during the initial region of the swerve maneuver (Mean Lat. Acc. Initial Maneuver). The T-scooter also attained higher––albeit not statistically significant––values than the other MMVs in all the other metrics. In contrast, all the handlebar-less devices––onewheel and three skateboards––attained generally low acceleration values in the first region of the maneuver. Furthermore, both the T-skateboard and the onewheel had the lowest lateral acceleration performances in the rebound region of the maneuver (Mean Lat. Acc. Rebound). In terms of jerk metrics, the high mean values observed from pedestrians exceeded the performance observed from all the devices. While not statistically significant, the jerk figures across the non-pedestrian devices supported the higher swerving performance of the T-scooter compared to all other MMVs.

*Table 8: Descriptive statistics (Mean ± Standard Deviation) and post-hoc test results for the MMV Type factor during swerving trials. Devices not connected by the same bracketed letters were significantly different (α=0.05).*

| | Mean Lat. Acc. Initial Maneuver [m/s$^2$] | Mean Lat. Acc. Rebound [m/s$^2$] | Mean Lat. Jerk Onset to Peak Acc. [m/s$^3$] | Mean Lat. Jerk Peak Acc. to Rebound Peak [m/s$^3$] | Mean Lat. Jerk Rebound Peak to End [m/s$^3$] |
|---|---|---|---|---|---|
| E-bicycle | 1.2±0.4 [C] | 0.7±0.3 [A,B] | 5.1±2.9 [B] | 3.9±2.1 [B] | 1.4±1.4 [B] |
| T-bicycle | 1.4±0.3 [B,C] | 0.6±0.3 [A,B] | 6.4±3.9 [B] | 4.5±2.1 [B] | 1.5±1.3 [B] |
| E-scooter | 1.5±0.6 [B,C] | 0.6±0.4 [A,B] | 7.9±8 [B] | 5.5±4.3 [B] | 1.4±1.9 [B] |
| T-scooter | 1.8±0.7 [A,B] | 0.6±0.3 [A,B] | 11.6±9.3 [B] | 6.7±3.7 [B] | 1.9±1.7 [B] |
| E-skateboard | 1.1±0.4 [C] | 0.6±0.3 [A,B] | 4.4±4.8 [B] | 4.2±2.5 [B] | 1.5±1.3 [B] |
| T-skateboard | 1.1±0.5 [C] | 0.5±0.2 [B] | 8.1±13.5 [B] | 4.3±2.6 [B] | 1.5±1.7 [B] |
| T-longboard | 1.0±0.4 [C] | 0.5±0.3 [A,B] | 5.8±6.8 [B] | 4.4±3.3 [B] | 1.2±0.9 [B] |
| Onewheel | 1.2±0.3 [C] | 0.5±0.2 [B] | 4.5±2.8 [B] | 3.8±1.4 [B] | 1.2±1.1 [B] |
| Pedestrian | 2.1±0.8 [A] | 0.8±0.4 [A] | 20.9±20.1 [A] | 28.5±26.8 [A] | 7.4±8.1 [A] |

The MMVs' swerving capabilities were significantly influenced by the different power sources (**Table 9**). Traditional bicycles and scooters exhibited higher lateral acceleration and jerk, to some extent, than their electric counterparts, opposite the direction observed within the braking analysis results (**Table 5**). In contrast, the E-skateboard attained significantly higher lateral acceleration than its traditional counterparts in the second phase of the maneuver (Mean Lat. Acc. Rebound). This difference, however, was not sizeable from a practical standpoint (**Table 8**).

*Table 9. Post-hoc test results for the MMV Power Source factor during swerving trials.*

|  | Significant Comparisons |
|---|---|
| Mean Lat. Acc. Initial Maneuver | T-bicycle > E-bicycle *<br>T-scooter > E-scooter * |
| Mean Lat. Acc. Rebound | E-skateboard > [T-skateboard, T-longboard] ** |
| Mean Lat. Jerk Onset to Peak Acc. | T-bicycle > E-bicycle **<br>T-scooter > E-scooter * |
| Mean Lat. Jerk Peak Acc. to Rebound Peak | T-bicycle > E-bicycle * |
| Mean Lat. Jerk Rebound Peak to Maneuver End | N.S. |
| ***α=0.01; **α=0.05; *α=0.10 | |

### 3.2.2. *Minimum radius of curvature*

The findings pertaining to the minimum radius of curvature obtained by each MMV showed that the onewheel exhibited the smallest radius of curvature, whereas the e-skateboard demonstrated the most extensive (**Table 10**). Speeds to obtain these minimum radii ranged between 1.2 and 1.7 m/s.

*Table 10. MMV minimum radius of curvature results.*

| MMV | Minimum radius [m] | Mean speed [m/s] |
|---|---|---|
| E-bicycle | 1.2 | 1.7 |
| T-bicycle | 0.9 | 1.3 |
| E-scooter | 1.2 | 1.5 |
| T-scooter | 0.6 | 1.4 |
| E-skateboard | 2.1 | 1.7 |
| T-skateboard | 1.2 | 1.4 |
| T-longboard | 1.5 | 1.6 |
| Onewheel | 0.3 | 1.2 |

## 4. Discussion

This investigation aimed to conduct an on-road experiment meant to gather kinematic data that effectively characterizes familiar MMV-rider performance capabilities within distinct categories of commonly used electric MMVs (e-MMVs), their conventional counterparts, and pedestrians. Data was acquired under controlled conditions involving emergency braking and swerving maneuvers. Our initial hypotheses were that the different MMV types and MMV propulsion systems would influence the device longitudinal and lateral evasive maneuvers, and that a piecewise linear model could be used to model MMV braking trajectories.

## 4.1. MMV types

### 4.1.1. *Braking*

In emergency braking trials, pedestrians displayed near-average deceleration values, yet registered the highest jerks among all the MMVs. This discrepancy is likely an artifact of the calculation of jerk used, which leveraged the peak deceleration. Pedestrians not only exhibited multiple peaks during their deceleration, but also exhibited sharp peaks due to the high-impulse contact between the pedestrian's shoe and the ground, which also generated higher friction than the MMV tire-road interface.

The exceptional bicycles and e-scooters mean deceleration, achieved with minimal jerk and a single braking ramp, suggests that the inherent design of bicycles and e-scooters contributes to rider stability during braking. While these findings align somewhat with previous observations (Garman et al. 2020, Lee et al. 2020, Li et al., 2023, Vetturi et al., 2023, Dozza et al., 2022, 2023), we obtained comparable performance between bicycles and the e-scooter, in contrast with the higher bicycles' capabilities observed in Dozza et al. (2022 and 2023) while performing harsh maneuvers. This discrepancy is likely attributable to the lower test speeds we employed, a factor proven to influence braking deceleration values by Vetturi et al. (2023) for e-scooters. Furthermore, the observed variation in performance may underscore the impact of different bicycle models, as exemplified by Li et al. (2023) in their analysis of two distinct e-scooter models. To comprehensively investigate this factor, future experiments should delve into diverse bicycle geometries and dimensions to identify potential correlations with braking performance. Additionally, the different braking systems on the devices could also deeply influence the results, as exemplified by the poor performance of the t-scooter. By generating the stopping force through the dragging of a foot-operated cover against the rear wheel, participants attained deceleration and jerk values that were even lower than those observed for handlebar-less MMVs in this study. Similarly, within those handlebar-less MMVs, the lower onewheel jerk value compared to the skateboards could be related to the different methods used to engage the brakes: skateboard riders employed foot-ground dragging (other than the remote control in the case of e-skateboard), while onewheel riders shifted their center of mass backward. Likely, this braking engagement method facilitated their gradual application of the electric brake (resulting in low jerk) but allowed onewheel riders to reach deceleration values higher than all the other skateboards. The numerical results of skateboards and onewheel are aligned with those obtained by Dozza et al. (2023) in examining the Segway, another frequently employed handlebar-less device. Their study design, however, did not control for participant experience, resulting in losing 50% of the Segway kinematic data, as participants were not able to complete the maneuvers. In our investigation, we specifically recruited participants familiar with these devices, expanding the applicability of their findings and thereby validating the generally inferior braking capabilities inherent in handlebar-less devices compared to bicycles and scooters.

### 4.1.2. *Swerving*

In lateral maneuvering, the pedestrians, who were not constrained by any vehicle structure, were able to execute more aggressive swerving maneuvers than MMV riders. This enhanced swerving capability underscores the unique advantages that pedestrians possess when navigating complex environments, and sets them apart from all the other examined subjects.

While the enhanced swerving performance of t-bicycles and scooters over handlebar-less devices were expected due to their intrinsic stability, the e-bicycle surprisingly exhibited one of the lowest lateral acceleration performances. This outcome may be attributed to its increased weight compared to other devices, a factor that could also help explain the excellent swerving capabilities demonstrated by the t-scooter, the lighter of the handlebar-equipped MMVs. Moreover, the notably lower positioning of the t-scooter handlebar in comparison to its electric counterpart may have facilitated participants in effectively applying steering torques. It is also noteworthy that studies by Dozza et al. (2022) and Garman et al.

(2020), involving controlled slalom maneuvers with e-scooters and bicycles, yielded significantly lower lateral mean acceleration values than those observed here. Instructing riders to swerve as hard as possible without knowing the direction and the location of the beginning of the maneuver, we closely simulated emergency steering maneuvers, ultimately establishing new limits for these devices' swerving capabilities.

Not surprisingly, the handlebar-less devices attained the lowest swerving performance observed, suggesting some degree of difficulty in completing these maneuvers with such devices. Similar to the braking results, the onewheel jerk was lower than the skateboard jerks for almost all the maneuver regions, confirming the inherent instability of this device. The four-wheel set up of the skateboards likely provided riders with slightly more stability, allowing them to quickly adjust their trajectories through body movements. However, compared to skateboard devices, the onewheel exhibited a significantly smaller minimum radius of curvature, highlighting its exceptional ability to execute tight swerving maneuvers. This could be attributed to the onewheel's compact design and single-point rotation mechanism, allowing it to pivot with minimal resistance. These findings suggest that the onewheel devices may exhibit less maneuverability at high speeds, but they could excel in navigating densely populated areas characterized by lower speeds.

### 4.2. MMV power sources

Previous research related to device power source has primarily focused on comparisons between traditional bicycles and e-bicycles (Dozza et al. 2023, Lee et al. 2023, Kovacsova et al. 2023), leaving a gap in understanding how different propulsion systems could impact the kinematic performance of other MMV types. In our study, we extended that scope by exploring the influence of different propulsion systems on braking and swerving capabilities across a wider range of MMVs.

In terms of braking metrics, the performance of the e-bicycle and e-scooter exceeded the performance of their traditional counterparts, validating results of previous research comparing traditional and electric bicycles in longitudinal maneuvers (Dozza et al., 2023; Li et al., 2023). This higher braking efficiency of e-MMVs could be partially related to the addition of electric (i.e., regenerative) braking to the conventional friction-based braking mechanism across all the e-MMVs. Further investigation should explore the relationship between specific electric device braking parameters and overall braking performance to establish optimal electric braking parameters for these devices.

Opposite results were obtained in swerving trials, with the traditional bicycle and scooter showing higher lateral acceleration and jerk than their electric counterparts. The decreased weight of traditional devices–– resulting from the absence of batteries and electric motors––could have potentially enhanced their maneuverability and responsiveness. This trend was not observed for the skateboard devices, where the e-skateboard had higher lateral acceleration in the rebound region of the maneuvers compared to traditional versions. Unlike bicycles and scooters, the increased weight of e-skateboards may not significantly impact rider swerving capability, given the inherently low center of mass of this device. The generally greater maneuverability of the traditional devices could also explain the reduced radius of curvature they were able to reach compared to their electric counterparts.

### 4.3. Maneuver limits and model limitations

#### 4.3.1. *Braking limits*

The experimental setup in this investigation ensures that the resulting data represents braking limits that can inform collision avoidance behaviors: participants were instructed to brake as much as possible while maintaining MMV control. The recruitment criteria selected participants who were regular MMV users so that the data presented participants' true capabilities. The results validated prior findings that MMV types can impact the longitudinal maneuverability in terms of mean braking, peak braking, and braking jerk (Dozza et al., 2023; Lee et al., 2020). Such results depict some MMVs' kinematic characteristics; however, they are not enough to construct a realistic trajectory with sustained initial braking, jerk, and final braking strength. To fill in that gap, this study applied piecewise linear models (Markkula et al., 2016) and produced parameterized functions to create braking profiles and trajectories that are essential in traffic event reconstructions and simulations. Crash reconstruction studies (Bareiss et al., 2019), and more recent simulations for autonomous vehicle (AV) collision avoidance capability testing, have been developed with logged vehicle trajectories (Kusano et al., 2022), which are limited in data quantity and coverage. The modeling results of this study can support the creation of additional scenarios, especially rare interactions with MMVs to test crash probabilities and AV safety performances. These derived braking limits also provide empirical data for making reasonable assumptions for safety envelope calculations (Singh et al., 2023) and safety-related models for driving behaviors (IEEE, 2022), as well as motion planning for AVs (Nayakanti et al., 2023). The piecewise linear models also revealed that the single brake ramp assumption (Markkula et al., 2016) does not apply to certain MMVs (specifically skateboards), and to pedestrians, due to fundamental differences in their braking or slowing down mechanisms. Further development of applicable modeling techniques for these scenarios remains as future work.

### 4.3.2. *Swerving limits*

Swerving trials were also designed to mimic emergency situations while maintaining MMV control, which provided a distribution of MMV users' evasive swerving capabilities via maximum lateral acceleration and jerk metrics. Additionally, the minimum curvatures associated with each MMV are novel contributions to the field. Like braking limits, empirically derived swerving can be used to make reasonable assumptions in safety envelope calculations and driving behavior models. For example, the pure pursuit model (Coulter, 1992), one of the most used swerving trajectory planning models, requires limits on agents' swerving capabilities when simulating trajectories. This is especially important for AVs, as they are mostly deployed in dense urban areas where interactions with MMVs are common (NACTO, 2020), yet naturalistic data involving MMVs is still rare. The swerving limits facilitate the generation of simulated MMV scenarios and the creation of realistic simulated MMV agents, and thus help AVs learn and improve interactions with MMVs for safer deployment.

## 5. Conclusions

The present study aimed to assess the emergency braking and swerving performances of experienced riders operating e-MMVs and their traditional counterparts. Exploring the impact of different MMV types and power sources, we gained valuable insights into the distinct kinematic features exhibited during critical maneuvers. Data from this study may support the development of solutions (e.g., training, regulations, and geofencing) that contribute to the safe integration of these vehicles into the current transport system, in several ways:

- Firstly, these findings underscore the importance of vehicle-specific characteristics, such as handlebars availability, wheelbase dimension, braking system design, and weight distribution, in influencing the stability, agility, and maneuvering capabilities of different MMV types. Beyond identifying similarities and distinctions that can inform the development of new regulatory frameworks, the identified MMV-related differences could serve as a valuable resource for designers and manufacturers seeking to comprehensively enhance the safety of micro-mobility transportation. Specifically, this study suggests the urgency of developing innovative solutions to improve braking capabilities for the handlebar-less devices, especially given the higher speeds achievable by their electric versions .
- Secondly, this study highlights the significant impact of different MMV propulsion systems in rapid and efficient emergency responses. The resulting substantial influence of weight considerations on the device agility and responsiveness suggest further research to explore innovative weight distribution schemes, aiming to specifically improve the e-MMVs' lateral performance.

- Finally, this research provides a new and useful tool for automated systems developers, increasing their ability to accurately predict MMV trajectories. The piecewise model results and the summary statistics presented in this paper may support the development and evaluation of intelligent systems and connected AVs by helping these systems and vehicles predict MMV riders' evasive performance in critical scenarios. While further research will still be required to model MMVs with fundamentally different braking or steering mechanisms from those assessed here, the data generated by this investigation fills in some sizeable gaps in the current literature.

## 6. Acknowledgements

The authors gratefully acknowledge the contributions of the additional experimenters who executed the data collection for this study, the developers who designed and implemented the instrumentation utilized, and the participants who provided their time and effort throughout the experimental trials. This research was funded by Waymo, LLC. The findings and conclusions of this paper are those of the authors and do not necessarily represent the views of the Virginia Tech Transportation Institute or Waymo LLC.

programme in Vienna: a retrospective multicentre study. *Archives of Orthopaedic and Trauma Surgery*, *141*(7), 1207–1213. https://doi.org/10.1007/s00402-020-03589-y

NACTO. (2020). *Shared micromobility in the US: 2019*.

Nayakanti, N., Al-Rfou, R., Zhou, A., Goel, K., Refaat, K. S., & Sapp, B. (2023). Wayformer: Motion Forecasting via Simple & Efficient Attention Networks. *2023 IEEE International Conference on Robotics and Automation (ICRA)*, 2980–2987. https://doi.org/10.1109/ICRA48891.2023.10160609

Pashkevich, A., Burghardt, T. E., Puławska-Obiedowska, S., & Šucha, M. (2022). Visual attention and speeds of pedestrians, cyclists, and electric scooter riders when using shared road – a field eye tracker experiment. *Case Studies on Transport Policy*, *10*(1), 549–558. https://doi.org/10.1016/j.cstp.2022.01.015

Petzoldt, T., Schleinitz, K., Krems, J. F., & Gehlert, T. (2017). Drivers' gap acceptance in front of approaching bicycles – Effects of bicycle speed and bicycle type. *Safety Science*, *92*, 283–289. https://doi.org/10.1016/j.ssci.2015.07.021

SAE International. (2019). *SAE J3194*. https://www.sae.org/standards/content/j3194_201911/

Schepers, J. P., Fishman, E., Den Hertog, P., Wolt, K. K., & Schwab, A. L. (2014). The safety of electrically assisted bicycles compared to classic bicycles. *Accident Analysis and Prevention*, *73*, 174–180. https://doi.org/10.1016/j.aap.2014.09.010

Schleinitz, K., Petzoldt, T., Franke-Bartholdt, L., Krems, J., & Gehlert, T. (2017). The German Naturalistic Cycling Study – Comparing cycling speed of riders of different e-bikes and conventional bicycles. *Safety Science*, *92*, 290–297. https://doi.org/10.1016/j.ssci.2015.07.027

Shah, N. R., Aryal, S., Wen, Y., & Cherry, C. R. (2021). Comparison of motor vehicle-involved e-scooter and bicycle crashes using standardized crash typology. *Journal of Safety Research*, *77*, 217–228. https://doi.org/10.1016/j.jsr.2021.03.005

Singh, H., Weng, B., Rao, S. J., & Elsasser, D. (2023). *A Diversity Analysis of Safety Metrics Comparing Vehicle Performance in the Lead-Vehicle Interaction Regime*. http://arxiv.org/abs/2306.14657

Smith, S. C., & Schwieterman, J. P. (2018). *E-Scooter Scenarios: Evaluating the Potential Mobility Benefits of Shared Dockless Scooters in Chicago*. https://www.researchgate.net/publication/330093998

Stigson, H., Malakuti, I., & Klingegård, M. (2021). Electric scooters accidents: Analyses of two Swedish accident data sets. *Accident Analysis and Prevention*, *163*. https://doi.org/10.1016/j.aap.2021.106466

Tark, J. (2020). *Micromobility Products-Related Deaths, Injuries, and Hazard Patterns: 2017 2019*.

Todd, J., Krauss, D., Zimmermann, J., & Dunning, A. (2019). Behavior of electric scooter operators in naturalistic environments. *SAE Technical Papers*, *2019-April*(April). https://doi.org/10.4271/2019-01-1007

Trivedi, B., Kesterke, M. J., Bhattacharjee, R., Weber, W., Mynar, K., & Reddy, L. V. (2019). Craniofacial Injuries Seen With the Introduction of Bicycle-Share Electric Scooters in an Urban Setting. *Journal of Oral and Maxillofacial Surgery*, *77*(11), 2292–2297. https://doi.org/10.1016/j.joms.2019.07.014

Trivedi, T. K., Liu, C., Antonio, A. L. M., Wheaton, N., Kreger, V., Yap, A., Schriger, D., & Elmore, J. G. (2019). Injuries Associated With Standing Electric Scooter Use. *JAMA Network Open*, *2*(1), e187381. https://doi.org/10.1001/jamanetworkopen.2018.738120